# Stabilization of STEP electrolyses in lithium-free molten carbonates

Stuart Licht*
*Department of Chemistry, George Washington University, Washington, DC 20052, USA*

This communication reports on effective electrolyses in lithium-free molten carbonates. Processes that utilize solar thermal energy to drive efficient electrolyses are termed <u>S</u>olar <u>T</u>hermal <u>E</u>lectrochemical <u>P</u>rocesses (STEP). Lithium-free molten carbonates, such as a sodium-potassium carbonate eutectic using an iridium anode, or a calcium-sodium-potassium carbonate eutectic using a nickel anode, can provide an effective medium for STEP electrolyses. Such electrolyses are useful in STEP carbon capture, and the production of staples including STEP fuel, iron, and cement.

Recently, a series of carbon dioxide splitting and electrochemical syntheses in molten carbonates were introduced.[1-5] Unlike, the traditional synthesis of lime (from limestone) or iron (from hematite or magnetite) these staples are produced without emission of $CO_2$. Additionally in molten carbontaes, carbon dioxide can be directly split to oxygen and carbon (at temperatures to ∼800°C), or to carbon monoxide (at temperature above 800C°). Each of these electrolyses is endothermic, and requires less energy at higher temperature. Processes that utilize solar thermal energy to provide this energy and drive efficient electrolyses are termed <u>S</u>olar <u>T</u>hermal <u>E</u>lectrochemical <u>P</u>rocesses (STEP). [6-7]

Large natural reserves of lithium, sodium and potassium carbonates, $Li_2CO_3$, $Na_2CO_3$ and $K_2CO_3$ exist. $Li_2CO_3$ melts are significantly more conductive than $Na_2CO_3$ or $K_2CO_3$ melts.[5] This high conductivity of lithium carbonate sustains high rates of electrolytic synthetic production ($A/cm^2$) at low overpotential.[5] However, the sodium and potassium salts are more prevalent and therefore a lithium-free molten carbonate could present economic advantages.[5]

We have explored the general energetics of molten carbonate electrolysis,[1] and previously calculated the thermochemical energetics of lithium carbonate compared to sodium or potassium carbonate electrolyses as a function of temperature.[2] Here, effective experimental electrolysis is demonstrated in such lithium-free carbonate melts.

$Li_2CO_3$ has a lower melting point (mp 723°C), than $Na_2CO_3$ (mp 851°C) or $K_2CO_3$ (mp 891°C), but a mix of the salts has lower melting point. We have previously explored effective electrolyses in both the pure $Li_2CO_3$ melt, and a $Li_{0.90}Na_{0.62}K_{0.48}CO_3$ melt. Mixed alkali carbonate melting points can be low, including 399°C for this eutectic $Li_{0.90}Na_{0.62}K_{0.48}CO_3$ mix, and 695°C for the $Na_{1.23}K_{0.77}CO_3$ eutectic salts. The addition of calcium carbonate can decrease the melting point of a carbonate mix. $CaCO_3$, as aragonite, decomposes at 825 °C, and as calcite melts at 1339 °C.[8] A variety of molten carbonates have been characterized with, and without, added calcium carbonates.[9-]

At 750°C in molten, pure lithium carbonate ($Li_2CO_3$, Alfa Aeasar, 99%) electrolysis was conducted with a nickel anode as the oxygen electrode (pure Ni 200 McMaster 9707K59) and a coiled steel wire (14 gauge) as the cathode. Experimental details of this and related molten lithium carbonate cells have been extensively detailed.[3,6] A thin nickel oxide overlayer forms on the anode, which is subsequently highly stable towards oxygen evolution.[3] The electrolysis evolves oxygen from the anode and forms a thick black graphite layer on the cathode. The carbon is readily removed from the cathode, by extracting, cooling and uncoiling the cathode wire (and the carbon then falls off the wire). The cathode product is characterized by x-ray powder diffraction (XRD powder diffraction data were collected on a Rigaku Miniflex diffractometer) and analyzed as graphite with the Jade software package.[14] The cathode wire can be used repeatedly, and does not exhibit signs of corrosion and remains the same diameter during repeated, and/or extended, electrolyses. The lithium 750°C molten carbonate cell sustains electrolysis potentials of less then 2 V at anodic current densities in excess of 100 mA cm², and this sustained electrolysis potential decreases to less than 1.0 V when high concentrations of $Li_2O$, such as 3.7 molal, are added to the electrolyte.[3] When calcium carbonate is added to the molten lithium carbonate, the electrolysis products are not only oxygen and reduced carbon, but also lime (calcium oxide) in accord with:

below 800°C: 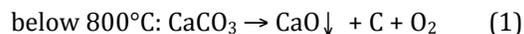  (1)

above 800°C: 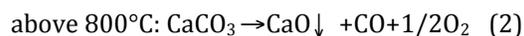  (2)

CaO is highly insoluble in molten $Li_2CO_3$ more dense, and precipitates during the electrolysis. As highly soluble limestone ($CaCO_3$) is continuously fed into the chamber during the electrolysis, lime, carbon and oxygen are produced and the electrolyte is unchanged.[6] Unlike calcium oxide, lithium oxide, $Li_2O$, is highly soluble in lithium carbonate. The solubility increases from 9 molal $Li_2O$ in 750°C $Li_2CO_3$ to over 1 molal $Li_2O$ at 950°C. During electrolysis in $Li_2CO_3$, $Li_2O$ does not precipitate, but reacts with $CO_2$ to continuously reform the $Li_2CO_3$ and leave the electrolyte unchanged, such as in the net of equations 3 and 4:

below 800°C: $Li_2CO_3 \rightarrow Li_2O + C + O_2$ (3)

$Li_2O + CO_2 \rightarrow Li_2CO_3$ (4)

net: $CO_2 \rightarrow C + O_2$ (5)

The carbon dioxide electrolysis potentials further decrease to less than 1.2 V (without added $Li_2O$) and less than 0.9 V (with added $Li_2O$) when iron, rather than carbon becomes the electrolysis product at the cathode (that is when 3.5m ferric oxide is also added to the molten electrolyte). Cathodic overpotentials increase the iron formation electrolysis potential to above 1V, but higher temperature, or higher concentration of ferric oxide decrease the electrolysis potential (to below 0.6 V for a 10 m Fe(III) dissolved in molten $Li_2CO_3$ electrolyte at 950°C.[3] These observe experimental electrolysis potential trends are generally consistent with our thermodynamic electrolysis potential trends as calculated from the free energy of formation of the constituents. Hence the calculated electrolysis potentials of the iron oxide splitting are less than that of the carbon dioxide splitting, Both decrease with increasing temperature, and the Nernst concentration corrected iron oxide splitting potential decreases with increasing ferric concentration.[1,3,15,16] For the STEP iron process, we had previously determined that $Fe_2O_3$ dissolves in lithium carbonate as the lithiated salt via:[2]

$Fe_2O_3 + Li_2O \rightarrow 2LiFeO_2$ (6)

In the process of electrolysis to form iron metal, $Li_2O$ is liberated:

$2LiFeO_2 \rightarrow 2Fe + Li_2O + 3/2O_2$ (7)

The liberated $Li_2O$ provides a path for the continued dissolution of $Fe_2O_3$, and overall the iron electrolysis process occurs as the combination of eqs 6 and 7:

net: $Fe_2O_3 \rightarrow 2Fe + 3/2O_2$ (8)

We have also demonstrated that stable carbon, calcium oxide, and iron production also occurs in a diminished lithium concentration carbonate melt. In these cases, a significant fraction of the lithium carbonate is replaced by sodium and potassium carbonates. These electrolyses have been demonstrated in the $Li_{0.90}Na_{0.62}K_{0.48}CO_3$ melt over a wide range of temperature and concentration conditions.[5,6] Silicates and aluminates occur in common iron ore and limestone deposits. We have also demonstrated that stable electrolysis occurs in molten electrolytes in which a fraction of the lithium carbonate has been replaced by silicate or aluminate salts.[3]

In this communication, unlike the lithium carbonate melt, under similar conditions, but when the electrolysis is conducted instead in a $Na_{0.23}K_{0.77}CO_3$ melt (without lithium carbonate and without added oxides) stable carbon dioxide splitting is not observed. Carbon may form, but tends to falls from the cathode due to excessive corrosion. Specifically, a 30 cm² surface area nickel foil anode is used as the oxygen electrode (pure Ni 200 McMaster 9707K59) and a 7.0 cm² surface area cathode is prepared by coiling a 18.3 cm long, 1.2 mm diameter steel wire. Anode and cathode are immersed in 50g of molten $Na_{0.23}K_{0.77}CO_3$ at 750°C. Electrolysis is conducted at a constant current of 0.5A for five hours, requiring an electrolysis potential of 2.5±0.1 V. In addition to not exhibiting a carbon product, the cathode wire also exhibits signs of corrosion during the ($Na_{0.23}K_{0.77}CO_3$ electrolysis, decreasing in diameter from 1.22 to 1.04 mm, as measured subsequent to the five hour electrolysis). While highly stable in lithium carbonate electrolytes, nickel anodes tend to corrode into molten sodium and potassium carbonate mix, observable as a green coloration developing in the electrolyte during extended electrolyses. During the electrolysis, nickel oxide dissolves into the electrolyte as evidenced by x-ray powder diffraction of the electrolyte, subsequent to the electrolysis, which exhibits nickel oxide, and of the cathode product which contains nickel metal (presumably as reduced from the dissolved nickel oxide in the electrolyte).

We have shown that iridium is a stable anode material in lithium carbonate melts. Iridium as an oxygen electrode exhibits a similar, low overpotential, and sustains similar high (A/cm²) anodic current densities as those observed at nickel anodes.[3] In addition, we report here that iridium is more stable than nickel in a lithium-free $Na_{1.23}K_{0.77}CO_3$ melt. The prepared iridium anode had a surface area of 3.7 cm² and consists of a coiled 22.3 cm long, 0.53 mm diameter iridium wire (0.50 mm, 99.8%, Alfa Aeasar), and is immersed with the 7.0 cm² area coiled steel wire cathode. The anode exhibits no change in thickness at 750°C during the course of repeated five hour electrolyses at a constant current 0.5 A electrolysis, and the electrolysis potential is stable between 2.3 to 2.5 V during the electrolysis.

Added calcium salts can decrease metal solubility and corrosion in molten sodium and potassium salt electrolytes. In molten $CaCl_2$ it was previously found that NiO solubility decreased with up to 4 mol% CaO concentration, and Ni coated with NiO had much higher stability during anodic polarization.[17] Molten carbonate electrolytic synthesis operates in the reverse mode of molten carbonate fuel cells (MCFC); where rather than fuel injection with electricity as a product, electrical energy is supplied and energetic chemical products are generated. MCFC systems have been studied in greater depth than carbonate electrolysis systems. Ni is a useful cell or electrode candidate material in MCFCs, but slowly degrades via a soluble nickel oxide overlayer. Cassir *et al.* report the of addition of 10% $CaCO_3$ to 650 °C $Li_{1.04}Na_{0.96}CaCO_3$ is useful to decrease the solubility of NiO from 150 to 100 μmolal in the carbonate mix.[18]

A more cost effective solution to the corrosivity of the sodium-potassium STEP carbonate melt (than the use of iridium electrodes) is found by the addition of calcium carbonate or barium salts to the sodium-potassium, lithium-free, carbonate melt (the addition of calcium carbonate is demonstrated here). The addition of calcium carbonate can decrease the melting point of a carbonate mix. The sodium/lithium carbonate mix, $Li_{1.07}Na_{0.93}CO_3$, has a melting point of 499°C, but decreases to below 450°C if 2 to 10 mol% equimolar $CaCO_3$ and $BaCO_3$ is added.[19]

In addition to the sodium-potassium carbonate electrolytes, electrolyses are also conducted here in calcium-sodium-potassium electrolytes ranging up to a calcium fraction of $Ca_{0.27}Na_{0.70}K_{0.75}$. Electrodes used are presented in Figure 1. A nickel oxygen anode appears to be fully stable during extended (five hour) 0.5 A electrolyses at 750°C in this melt, using the 30 $cm^2$ nickel foil anode and a 7.0 $cm^2$ steel wire cathode, and the electrolysis proceeds at between 1.9 to 2.2V. Unlike the electrolyses conducted in the calcium free (sodium-potassium) carbonate melt, carbon forms and remains on the cathode during electrolysis, and the steel cathode remains the same diameter, as measured subsequent to the electrolysis. As shown subsequent to the electrolyses, in the cathode photographs at the bottom of Figure 1, electrolyses conducted in either $Ca_{0.16}Na_{1.03}K_{0.65}$ or $Ca_{0.27}Na_{0.70}K_{0.75}$ electrolytes exhibit a thick carbon product on the cathode, while this is not the case following electrolysis without calcium carbonate in $Na_{1.23}K_{0.77}CO_3$. The electrolysis potential and subsequent cathode product, during a repeat of the $Ca_{0.27}Na_{0.70}K_{0.75}$ electrolysis, but at a constant electrolysis current of 1A, rather than 0.5A, is presented in Figure 2.

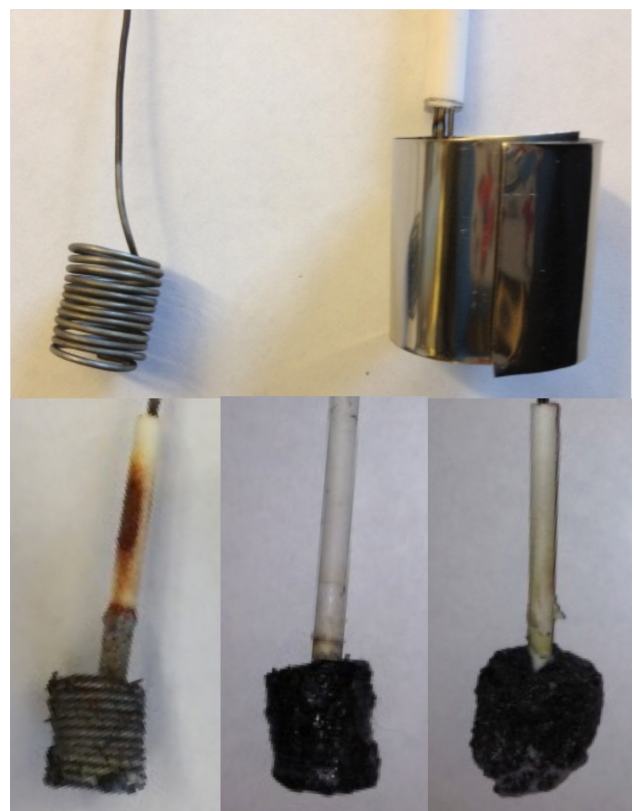

**Figure 1.** Top: Cathode (top left) and anode (top right) prior to 0.5 A, 5 hour lithium-free electrolyses at 750°C with increasing calcium carbonate concentration. The cathode is placed inside the anode, which are both immersed in the molten electrolyte. Bottom: Cathodes after electrolysis in lithium-free molten carbonates. Electrolytes used were respectively: $Na_{1.23}K_{0.77}CO_3$ (lower left cathode), $Ca_{0.16}Na_{1.03}K_{0.65}$ (lower middle cathode), and $Ca_{0.27}Na_{0.70}K_{0.75}$ (lower right cathode).

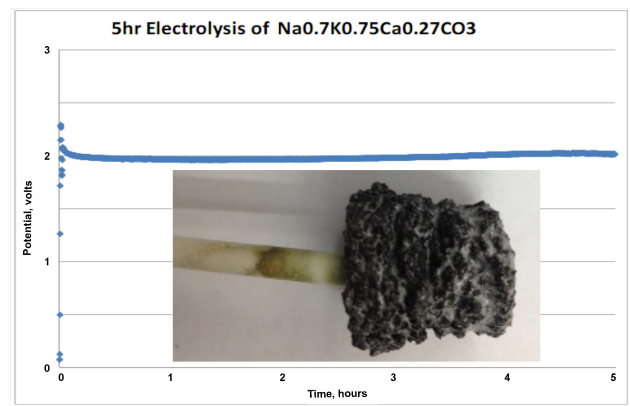

**Figure 2.** Time variation of the electrolysis potential during a five hour electrolysis at 1A in $Ca_{0.27}Na_{0.70}K_{0.75}$ at 750°C. Inset: cathode subsequent to the electrolysis.

The observed increased the stability of the electrolysis in the lithium free (sodium-potassium) carbonate electrolyte containing calcium carbonate may be related to the low solubility of calcium oxide, compared to sodium oxide in these electrolytes. Whereas we find that sodium oxide, $Na_2O$, is fully miscible in molten sodium-potassium carbonates. Calcium oxide is nearly insoluble (solubility less than 0.1 molal) from 750 to 950°C, in $Na_{1.23}K_{0.77}CO_3$ or $Ca_{0.27}Na_{0.70}K_{0.75}$. In general the strength of an oxide in melts to donate an electron pair (Lewis basicity) decreases in the order: $K_2O > Na_2O > Li_2O > BaO > CaO > MgO > Fe_2O_3 > Al_2O_3 > TiO_2 > B_2O_3 > SiO_2 > P_2O_5$.[20] The lower basicity, combined with the lower concentration, of the oxide in the calcium containing carbonate electrolyte contribute to the observed decrease in corrosivity and improved electrolysis.

Acknowledgement. The United States National Science Foundation and the George Washington University provided partial funding for this study. Doctoral graduate student Ulyana Cubeta, who is no longer pursuing this project, participated in the collection of this data. These results will be included in a broader publication related to this topic.

---

*To whom correspondence should be addressed. Email: slicht@gwu.edu